\documentclass[aps,pre,twocolumn,showpacs,superscriptaddress]{revtex4}
\usepackage[latin1]{inputenc}
\usepackage[english]{babel}
\usepackage{bm,graphicx,graphics,amsmath,amssymb,epsfig}
\usepackage{txfonts}

\usepackage[colorlinks,citecolor=blue,linkcolor=Fuchsia]{hyperref}
\usepackage[usenames,dvipsnames,svgnames]{xcolor}

\def \be {\begin{equation}}
\def \ee {\end{equation}}
\def \laplace {\triangle}

\def \average#1{\left\langle #1 \right\rangle}

\def \graffb#1{\left\{ #1 \right\}}

\begin{document}

\title{Energy and magnetisation transport in non-equilibrium macrospin systems}
\author{Simone Borlenghi} 
\affiliation{Department of Physics and Astronomy, Uppsala University, Box 516, SE-75120 Uppsala, Sweden.}
\affiliation{Department of Materials and Nanophysics,  School of Information and Communication Technology, \\Electrum 229, Royal Institute of Technology, SE-16440 Kista, Sweden.}
\author{Stefano Iubini}
\affiliation{Centre de Biophysique Moléculaire (CBM), CNRS-UPR 4301 Rue Charles Sadron, F-45071 Orléans, France}
\author{Stefano Lepri}
\affiliation{Consiglio Nazionale delle Ricerche, Istituto dei Sistemi Complessi, Via Madonna del Piano 10 I-50019 Sesto Fiorentino, Italy.} 
\affiliation{Istituto Nazionale di Fisica Nucleare, Sezione di Firenze, 
via G. Sansone 1, I-50019 Sesto Fiorentino, Italy}
\author{Jonathan Chico}
\affiliation{Department of Physics and Astronomy, Uppsala University, Box 516, SE-75120 Uppsala, Sweden.}
\author{Lars Bergqvist} 
\affiliation{Department of Materials and Nanophysics,  School of Information and Communication Technology, \\Electrum 229, Royal Institute of Technology, SE-16440 Kista, Sweden.}
\author{Anna Delin}
\affiliation{Department of Materials and Nanophysics,  School of Information and Communication Technology, \\Electrum 229, Royal Institute of Technology, SE-16440 Kista, Sweden.}
\affiliation{Department of Physics and Astronomy, Uppsala University, Box 516, SE-75120 Uppsala, Sweden.}
\author{Jonas Fransson}
\affiliation{Department of Physics and Astronomy, Uppsala University, Box 516, SE-75120 Uppsala, Sweden.}
\date{\today}

\begin{abstract}
We investigate numerically the magnetisation dynamics of an array of nano-disks interacting 
through the magneto-dipolar coupling. In the presence of a temperature gradient, the chain reaches a non-equilibrium steady state where energy and magnetisation currents propagate. This effect 
can be described as the flow of energy and particle currents in an off-equilibrium discrete nonlinear Schr\"odinger (DNLS) equation. This model makes transparent the 
transport properties of the system and allows for a precise definition of temperature and chemical potential for a precessing spin. The present study proposes a novel setup for the spin-Seebeck effect, and shows that its qualitative features
can be captured by a general oscillator-chain model
\end{abstract}

\maketitle 

\section{Introduction}

Off-equilibrium dynamics of classical and quantum many-particle systems is 
is a wide topic, ranging from the theoretical 
foundations to the development of innovative ideas for nanoscale
thermal management.  
The nanoscale control of heat flows offer promising opportunities for novel energy harvesting devices, and constitutes a fertile terrain to with possible future applications
to nanotechnologies and effective energetic resources.

In this context, simple models of classical nonlinear oscillators have 
been investigated to gain a deeper understanding of heat transfer processes far 
from thermal equilibrium \cite{lepri03,basile08,DHARREV}. In the case of 
systems admitting two (or more) conserved quantities, coupled transport (for instance 
of energy and mass) is an issue of basic relevance in connection with thermoelectric phenomena~\cite{Saito2010} whereby temperature gradients can be employed
to generate electric currents. 

A related topic concerns coupled transport in magnetic systems. 
The recent discovery of the spin-Seebeck effect has opened the new field of 
spin-caloritronics \cite{uchida08,uchida10,bauer12}, where the transport properties of 
thermally driven spin systems is in focus. The propagation of spin wave (SW) current driven
by a thermal gradient has been extensively studied in systems of exchange coupled spins, 
using the well established micromagnetic formalism \cite{ohe11,hinzke11,ritzmann14,etesami14}.
In the context of statistical mechanics,  
although several studies of heat transport in classical Heisenberg spin chains exists \cite{Savin2005,Bagchi2012}, this topic has not been treated in detail up to now.

Here, we investigate transport in a novel setup, which consists of an array of 
magnetic nano-disks coupled through the magneto-dipolar interaction, and interacting
with different external reservoirs.  As a result, 
energy and magnetisation currents are carried by damped dipolar spin waves.

To obtain a better insight, we follow a simplified physical picture: one can think of a spin system as a chain of nonlinear oscillators. Intuitively, a thermal bath acts as a random force on some part of the chain, whose effect is the propagation of oscillations (spin waves) in the system, in the form of energy and magnetisation currents. More specifically, we will model the system as an open discrete nonlinear Schr\"odinger (DNLS) system that steadily exchanges energy with
external reservoirs, a setup that has been considered only recently 
\cite{Basko2011,iubini12,DeRoeck2013}. In particular, we will employ 
a Langevin thermostatting scheme whereby complex white noises and dissipative couplings
are added to the equation  \cite{iubini13,borlenghi14c}. Altogether, this guarantees that 
the chain reaches thermal equilibrium when put in contact with thermostats at the same 
temperature. As we will show, this approach has several advantages in terms 
of simplicity of description and it makes transparent the 
transport/thermodynamical properties of the micromagnetic system.

This paper is organised as follows. In Section II we review the equation of motion of the magnetisation for a macrospin interacting with a reservoir. 
In Section III we outline the derivation of the simplified oscillator model
and describe its relations with the micromagnetic equations. In Section IV we study  the
steady-state properties of a spin-chain made of ten disks coupled via
dipolar interaction through micromagnetic simulations. Using the formalism of the DNLS
equation, the currents and the effective spin temperature of the system are calculated. 
The results are compared with the direct simulation of the oscillator model in Section V. 
Finally, the conclusions of this work are contained in the last Section.

\section{Physical system}%
 
\begin{figure}[t]
\begin{center}
\includegraphics[width=8cm]{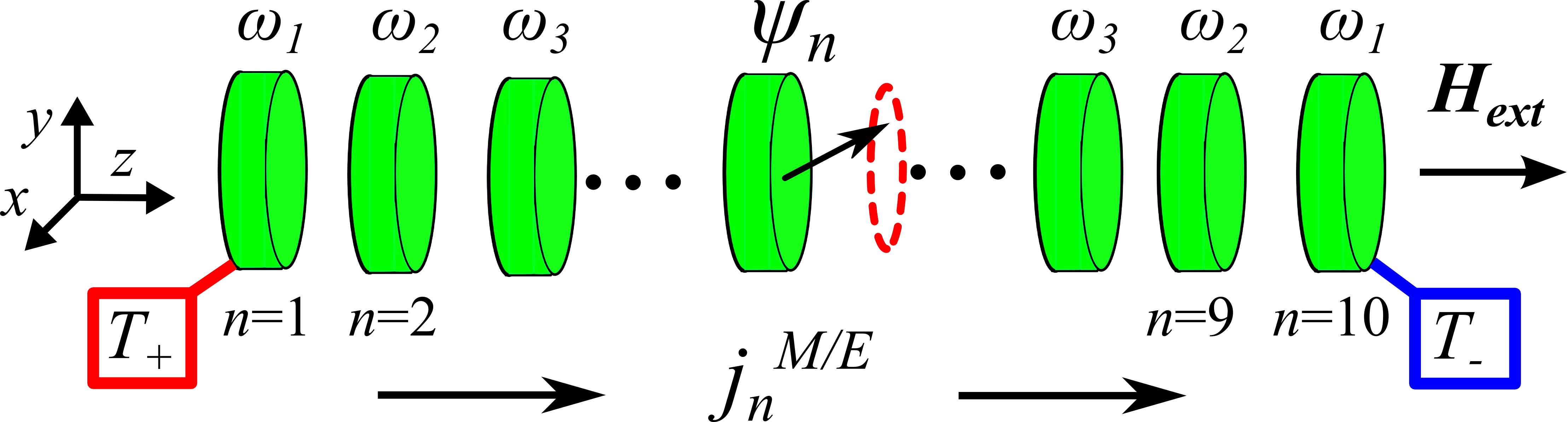}
\end{center}
\caption{Chain of disks coupled via dipolar interaction. Each disk behaves as a macrospin with precession frequency $\omega_n$. The first and last element of each chain are 
coupled to thermal reservoirs with temperatures $T_\pm$, that control the propagation of energy and magnetisation currents.}
\label{fig:figure1}
\end{figure}

The system studied here, shown in Fig.\ref{fig:figure1} consists of an array of $N=10$ identical nano-disks made of Permalloy (Py), coupled through the dipolar interaction.
The first and last disks are coupled to Langevin thermal baths with temperatures respectively $T_{\pm}$. Thermal fluctuations excite the SW modes of the
system. In the presence of a temperature difference $\Delta T=T_+-T_-$, the disk chain reaches a non-equilibrium stationary state where  two coupled currents, of energy and magnetisation, flow from the hot  ($T_+$) to the cold ($T_-$) reservoirs.

The dynamics is investigated by means of micromagnetic simulations.
The cornerstone of micromagnetism is the Landau-Lifschitz-Gilbert (LLG) equation of motion for a ferromagnet \cite{landau65,gilbert55,gurevich96}

\be\label{eq:LLG}
\dot{\bm M}= \gamma {\bm M}\times {\bm H}_{\rm{eff}} +\frac{\alpha}{M_s}{\bm M}\times{\dot{\bm M}},
\ee
which describes the precession of the local magnetisation vector $\bm{M}(\bm{r},t)$ around the effective field ${\bm H}_{\rm{eff}}$.
The first term of Eq.(\ref{eq:LLG}), proportional to the gyromagnetic ratio $\gamma$, accounts for the precession. The second term describes energy dissipation at a rate proportional to the dimensionless Gilbert damping
parameter $\alpha$. In the absence of an external driving field, such as thermal fluctuations, spin transfer torque or rf fields, the magnetisation eventually aligns with ${\bm H}_{\rm{eff}}$ \cite{gurevich96,slavin09}.
The saturation magnetisation $M_s$ is the norm of the magnetisation vector, which depends on the material and the geometry of the sample.
The thermodynamical properties of the system are described by the Gibbs free energy
\begin{align}
\label{eq:free}
\mathcal{F}=&\mu_0\int_V[A\laplace{\bm{M}}(\bm{r},t)-{\bm{M}}(\bm{r},t)\cdot{\bm{H}}_{\rm{ext}}
\nonumber\\
                 	&- \frac{1}{2}\bm{M}(\bm{r},t)\cdot\bm{H}_{\rm{dip}}]{\rm{d}}^3r,
\end{align}
$\mu_0$ being the vacuum magnetic permeability. In the present case, Eq.(\ref{eq:free}) contains contributions respectively from exchange energy, Zeeman interaction and dipolar interactions.
The effective field in Eq.(\ref{eq:LLG}) is given by the functional derivative $\bm{H}_{\rm{eff}}=-(1/\mu_0)\delta\mathcal{F}/\delta\bm{M}$ of the Gibbs free energy with respect to the magnetisation.
It is the sum of the following three terms:

\begin{subequations}
\begin{align}
\bm{H}_{\rm{exc}} = & -A\nabla \bm{M}({\bm{r}},t),\label{eq:exch}\\
\bm{H}_{\rm{ext}} = & H\hat{\bm{z}},\label{eq:ext}\\
\bm{H}_{\rm{dip}} = & -\frac{1}{4\pi}\int_V\frac{\rho_M(\bm{r}-\bm{r}^\prime)}{|\bm{r}-\bm{r}^\prime|^3}{\rm{d}}^3r^\prime
                             + \frac{1}{4\pi}\int_S \frac{\sigma_M(\bm{r}-\bm{r}^\prime)}{|\bm{r}-\bm{r}^\prime |^3}{\rm{d}}^2r^\prime. \label{eq:dip}
\end{align}
\end{subequations}
Eqs. (\ref{eq:exch}) and (\ref{eq:ext}) are respectively the exchange field with exchange constant $A$ and the external field with intensity $H$ along $\hat{\bm{z}}$.
The exchange field is the short range interaction responsible for the coherent precession of the magnetisation inside each disk, while the applied field defines the precession axis.
The dipolar stray field Eq.(\ref{eq:dip}) contains contributions from the volume charges $\rho_M=\nabla\cdot\bm{M}({\bm{r}^\prime},t)$ and the surface charges $\sigma_M=\bm{M}({\bm{r}^\prime},t)\cdot\hat{\bm{n}}$,
where $\hat{\bm{n}}$ denotes the normal to the surface of the sample at point $\bm{r}^\prime$. The dipolar field acts as a demagnetising field in each disk and adds a coupling between the disks. It is responsible for the nonlinearity of the LLG equation, Eq. (\ref{eq:LLG}).

Thermal fluctuations are introduced by adding to the effective field the stochastic term 

\be\label{eq:th}
{\bm H}_{\rm{th}}(\bm{r},t)=\sqrt{DT}(\eta_x,\eta_y,\eta_z),
\ee 
where $\eta_j(\bm{r},t)$, $j=(x,y,z)$, is a Gaussian random process with zero average and correlation $\average{\eta_j({\bm{r}},t)\eta_{j^\prime}({\bm{r}}^\prime,t^\prime)} =\delta_{jj^\prime}\delta(\bm{r}-\bm{r}^\prime)\delta(t-t')$. 
\cite{martinez07,grinstein03}. The term $T=T(\bm{r})$ denotes the local temperature of the underlying phonon bath, while the coupling strength with the bath is

\be\label{eq:diffusion}
D=\frac{2\alpha k_B}{\gamma \mu_0 V_M M_s}.
\ee\label{eq:diffconst}
Here $k_B$ is the Boltzmann constant and $V_M$ is the volume associated to the magnetic moment $\bm{M}(\bm{r})$. 

In our micromagnetics simulations, where the sample is represented by a finite element tetrahedral mesh, the LLG equation, Eq. (\ref{eq:LLG}), is solved numerically at each mesh node. 
The coordinate $\bm{r}$ is discretised and corresponds to the positions of the nodes, while $V_M$ corresponds the volume of each mesh elements.

\section{Coupled oscillator model}%

\begin{figure}[t]
\begin{center}
\includegraphics[width=3cm]{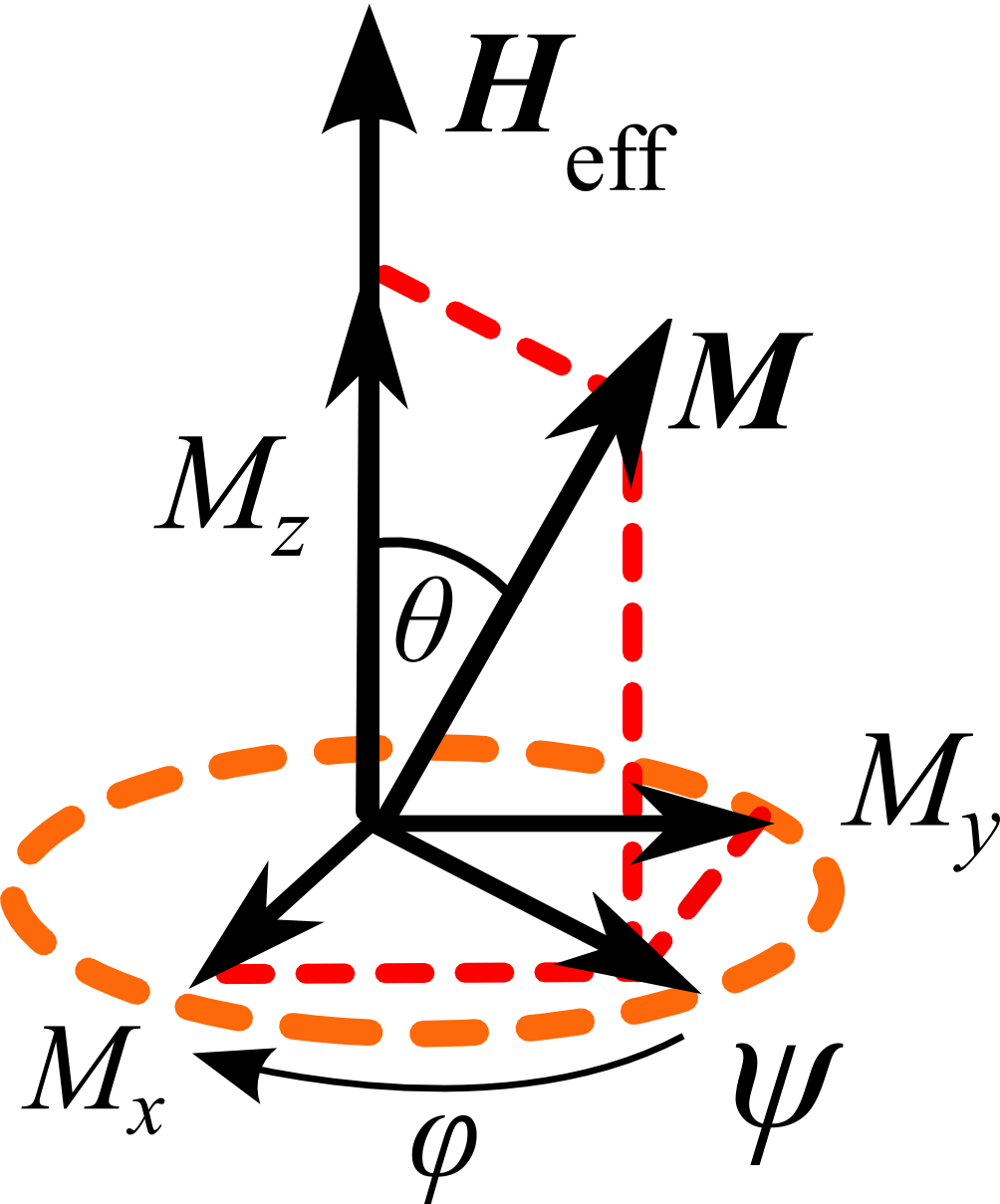}
\end{center}
\caption{Macrospin $\bm{M}$ precessing around the effective field $\bm{H}_{\rm{eff}}$, described in terms of the stereographic projection $\psi$. The phase $\phi$ describes the dynamics in the $x$-$y$ plane,
while the SW power $p=|\psi|^2$ is related to the polar angle $\theta$ (see text).}
\label{fig:figure2}
\end{figure}

The dynamics of the chain is conveniently described by the volume-averaged magnetisation inside the $n$th disk:

\be\label{eq:vol_average}
\bm{M}^n(t)=\frac{1}{V_n}\int_{V_n}\bm{M}(\bm{r}_n,t){\mbox{d}}^3r_n.
\ee
Due to the uniform precession of the magnetisation in each disk,  the system can be modeled as an assembly of coupled macrospins $\bm{M}_n(t)$.
With this approximation, Eq.(\ref{eq:LLG}) can be written in the form of an equation of motion for an ensemble of coupled nonlinear oscillators \cite{slavin09}

\begin{align}\label{eq:dnls}
i\dot{\psi}_n =& \omega_n(p_n)\psi_n-i\Gamma_n(p_n)\psi_n+\sum_{n^\prime}J_{nn^\prime}\psi_{n^\prime}
\nonumber\\&
	+\sqrt{D_n(p_n)T_n}\xi_n,
\end{align}
where the complex SW amplitudes are defined by 

\be\label{eq:psi}
\psi_n=\frac{M_x^n-iM_y^n}{\sqrt{2M_s(M_s+M_z^n)}}.
\ee
By writing Eq.(\ref{eq:psi}) as $\psi_n=\sqrt{p_n(t)}\mbox{e}^{i\phi_n(t)}$, one can see that 
$\phi_n$ describes the precession of $\bm{M}$ in the $x$-$y$ plane, while $p_n=|\psi_n|^2$ (referred to as the local SW power) is related to the polar angle
through $\theta_n=\arccos(1-p_n)$, see Fig.\ref{fig:figure2}.

The first two terms on the right hand side of Eq.(\ref{eq:dnls}) are respectively the nonlinear frequencies $\omega_n(p_n)$ and
damping rates $\Gamma_n(p_n)$ of the $n_{\rm{th}}$ disk. Both are proportional to the effective field $\gamma |\bm{H}_{\rm{eff}}^n\cdot{\hat{\bm{z}}}|$ acting on each disk.

For the small precession amplitudes considered here, nonlinear effects are taken into account pertubatively by expanding into powers of $p_n$ the frequencies and damping rates,
respectively as \cite{slavin09} $\omega_n(p_n)\approx\omega_n+\nu p_n$, $\Gamma_n(p_n)\approx\alpha\omega_n(1+Qp_n)$. Here $\omega_n$, $\nu$ and $Q$ are 
the coefficients of the expansion of $\omega_n(p_n)$ and $\Gamma_n(p_n)$  to the first order in $p_n$. 
The third term on the right hand side is the interlayer coupling $J_{nn^\prime}$. 
In general it is a complex quantity, and its phase is related to energy gain or dissipation \cite{borlenghi14c}.
In the following we will consider the simple case of a uniform nearest-neighbour interaction
that amounts to retain only terms containing  $\psi_{n\pm 1}$ in Eq.(\ref{eq:dnls}) 
and set $J_{nn^\prime}=J\delta_{n,n\pm1}$.
Note also that the imaginary part of $J_{nn'}$ and $\Gamma_n$ are 
related since they both stem from the dissipative term proportional to $\alpha$ in
the LLG equation, Eq. (\ref{eq:LLG}).
Both the nonlinearity and the coupling are due to the dipolar field Eq.(\ref{eq:dip}).
The parameters $(\omega_n,\nu,Q,J_{nn^\prime})$ can be calculated analytically only in some simple cases, but in general they must be inferred from micromagnetic simulations
\cite{slavin09,naletov11}.

We consider the case where the temperature is uniform within each disk.
In this case, the last term of Eq.(\ref{eq:dnls}) describes thermal fluctuations in terms
of the \emph{complex} Gaussian random 
variables $\xi_n=(\eta_x^n+i\eta_y^n)$ and the  
nonlinear diffusion constant $D_n(p_n)$.  In the linear regime, the latter equals 
$\gamma D$, where $D$ is the quantity defined in Eq.(\ref{eq:diffusion}).
In the nonlinear regime $D_n(p_n)$ depends on $\Gamma_n(p_n)$, $\omega_n(p_n)$ and $J_{nn'}$ 
and must be fixed consistently to satisfy the fluctuation-dissipation theorem. 
For the single oscillator this amounts to fix $D_n(p_n)\propto\Gamma_n(p_n)/\omega_n(p_n)$\cite{slavin09}. 
This is necessary to ensure that, for $T_n\equiv T$, the systems approaches a global 
canonical equilibrium at temperature $T$. Consistently with the small-amplitude limit,
we assume  $\Gamma_n(p_n)=\alpha\omega_n(p_n)$. As a consequence, the noise term in 
Eq.(\ref{eq:dnls}) becomes
purely additive and $D_n(p_n)$ reduces to a constant $D=\alpha$ in units with $k_B=1$.

Taking into account the above approximations, we simplify Eq.(\ref{eq:dnls}) into
\begin{align}
i \dot {\psi}_n =&(1+i\alpha)\left[-\nu |\psi_n|^2 \psi_n -\omega_n \psi_n 
-J(\psi_{n+1}+\psi_{n-1})\right]\nonumber \\ 
& +\sqrt{\alpha T_n} \, \xi_n(t) 
\quad.
\label{eq:ourdnls}
\end{align}
Upon defining the ``Hamiltonian"
\be\label{eq:hamiltonian}
\mathcal{H} = {\sum_n}[{\omega_n |\psi_n|^2 +\frac{\nu}{2} |\psi_n|^4
+J(\psi_n^*\psi_{n+1}+\psi_n\psi_{n+1}^*)}],
\ee
where $(\psi_n,i\psi_n^*)$ are canonically conjugate variables
satisfying the Hamilton equations  $\dot\psi_n=-\partial\mathcal H/\partial i\psi_n^*$,
Eq.(\ref{eq:ourdnls}) can be written more concisely in the form of a Langevin
equation with uniform bath coupling $\alpha$
$$
i\dot{\psi}_n=-(1+i\alpha)\frac{\partial\mathcal{H}}{\partial\psi_n^*} 
+\sqrt{\alpha T_n} \, \xi_n.  
$$ 

In the absence of coupling with the external baths $(\alpha=0)$, Eq.(\ref{eq:ourdnls}) is the well-known 
DNLS equation \cite{Eilbeck1985} which, at variance with
its continuum limit, is not integrable \cite{rumpf03}. Such equation 
describes a large class of conservative oscillating systems.
Some examples include transport in biomolecules,  
Bose-Einstein condensates in optical lattices, mechanical oscillators and photonics waveguides
\cite{eilbeck03,Kevrekidis}. The dependence of $\omega_n$ on the lattice 
sites introduces an heterogeneity that is also connected with the nonlinear
version of the Anderson tight-binding model \cite{Pikovsky2008,Kopidakis2008,Basko2011}.

In the non dissipative limit, the model admits a further constant of motion besides energy, namely, the total SW 
power $\mathcal{P}=\sum_n p_n$. As a consequence, the thermodynamic equilibrium phase-diagram
is two-dimensional, and each equilibrium state is determined by the energy density $\mathcal{H}/N$ and the SW 
density $\mathcal{P}/N$ ~\cite{Rasmussen2000}. 

In the nonequilibrium regime, the local fluxes of the 
conserved quantities are of special interest.
By computing the time derivatives of the SW power $p_n$ and of the 
local energy $\mathcal{H}_n$, one obtains the two
continuity equations \cite{slavin09,iubini13,borlenghi14a,borlenghi14b,borlenghi14c}
\begin{subequations}
\begin{align}
\dot{p}_n = & \mathcal{J}^M_n+j_{n+1}^M-j_n^M,\label{eq:consp}\\
\dot{\mathcal{H}}_n = &  \mathcal{J}^E_n +j_{n+1}^E-j_{n}^E,\label{eq:consh}
\end{align}
\end{subequations}
where
\begin{subequations}
\begin{align}
j_n^M = & 2J\,{\rm{Im}}[\psi_n^*(\psi_{n+1}-\psi_n)],\label{eq:mag}\\ 
j_n^E = & 2J\,{\rm{Re}}\left[\frac{\partial \mathcal H}{\partial i \psi_n}(\psi_{n+1}-\psi_n)\right].\label{eq:energy}
\end{align}
\end{subequations}
are the magnetisation and energy currents associated with the hamiltonian coupling
between neighboring oscillators. The remaining terms $\mathcal{J}^M_n$ and $\mathcal{J}^E_n$
account for the exchange with the reservoirs due to both fluctuations and damping. 
Their steady-state averages are computed using stochastic calculus, 
by evaluating the change of $p_n$ and $\mathcal{H}_n$ up to second order in the noise term
\cite{lepri03}. As a result, one gets 
\begin{equation} \label{eq:bfluxM}
\langle \mathcal{J}^M_n \rangle 
= 2 \alpha T_n -2\alpha\left\langle \left[ \omega_n(p_n)p_n+ 
J {\rm{Re}}\left(\psi_n^*(\psi_{n+1}+\psi_{n-1}\right)\right] \right\rangle \\
\end{equation}
A similar  (but more involved) expression holds for the energy fluxes.
As a preliminary test, we verified that for a generic nonequilibrium stationary state
the above definitions of fluxes satisfy the local flux balance expressed in Eqs.  (\ref{eq:consp}) and (\ref{eq:consh}). 

For a system which is driven out-of-equilibrium from its boundaries, the local temperature represents a 
useful observable for the characterisation of the stationary state.
Note in particular that the
quantity $T_n$ that appears in Eq.(\ref{eq:dnls}) specifies the temperature of the
phonon bath, which in general \emph{does not} correspond to the temperature of the system \cite{iubini12}. 
The definition of temperature for a system of interacting  magnetic moments is not straightforward, since the
model Hamiltonian  is non separable, and one cannot relate temperature to the average kinetic energy.
Within the DNLS formalism, one can use the general microcanonical definition of temperature for non separable
Hamiltonians with two conserved quantities \cite{franzosi11}.  
The general expression is nonlocal and rather involved, we refer to Refs.\cite{franzosi11,iubini12,iubini13} 
for details. 
Since we are interested in the limit of low temperatures and low amplitudes, we can follow the derivation
in Ref. \cite{iubini13} and introduce a simple approximation of the microcanonical temperature based on 
a mapping of the DNLS equation to a chain of nonlinear coupled rotators (XY model).
Accordingly, the temperature is approximated by  

\be\label{eq:Txy}
T_{XY}=f(\average{p_n})\left[\average{\dot{\phi}_n^2}-\average{\dot{\phi}_n}^2\right]
,
\ee
and  it acquires a simple interpretation of the phase-fluctuations  of the oscillator $\psi_n$.
The function $f(\average{p_n})$ is a rescaling factor that depends on the average local power $\average{p_n}$.
One can also show that 
for $\average{p_n}\ll 1$, $f(\average{p_n})=2\average{p_n}$.

\section{Micromagnetic simulations}\label{sec:MMag}%

The micromagnetic simulations were performed with the Nmag software \cite{fischbacher07}, using a tetrahedral finite
element mesh with maximum size of 3 nm, 
of the order of the Py exchange length, and an integration time step is 1 ps. The mesh was automatically generated by
the Netgen package \cite{schoeberl97}.

Each disk of the chain has thickness $t=3$ nm, radius $R=20$ nm, and an interlayer distance $d=3$ nm. The applied 
field $\bm{H}_{\rm{ext}}=1$T defines the precession axis of the magnetisation along $\hat{\bm{z}}$.
The exchange stiffness $A=1\times 10^{-11}$ J/m corresponds to that of Py, while the other micromagnetic parameters are
$M_{s}=0.94$ T/$\mu_0$, $\alpha=8\times 10^{-3}$ and $\gamma=1.873\times 10^{11}$ rad$\times$s$^{-1}$ $\times$T$^{-1}$. 

The output of the simulations consists of the ensemble of the magnetisation vectors $\graffb{\bm{M}^n({\bm{r}}_n,t)}$,
$n=1,...,10$. Each vector depends on the spacial coordinate ${\bm{r}}_n$ inside 
the $n_{\rm{th}}$ disk, and the collective magnetisation dynamics of each disk is given by the volume average Eq.(\ref{eq:vol_average})
\cite{slavin09,borlenghi14b}

\begin{figure}[t]
\begin{center}
\includegraphics[width=7cm]{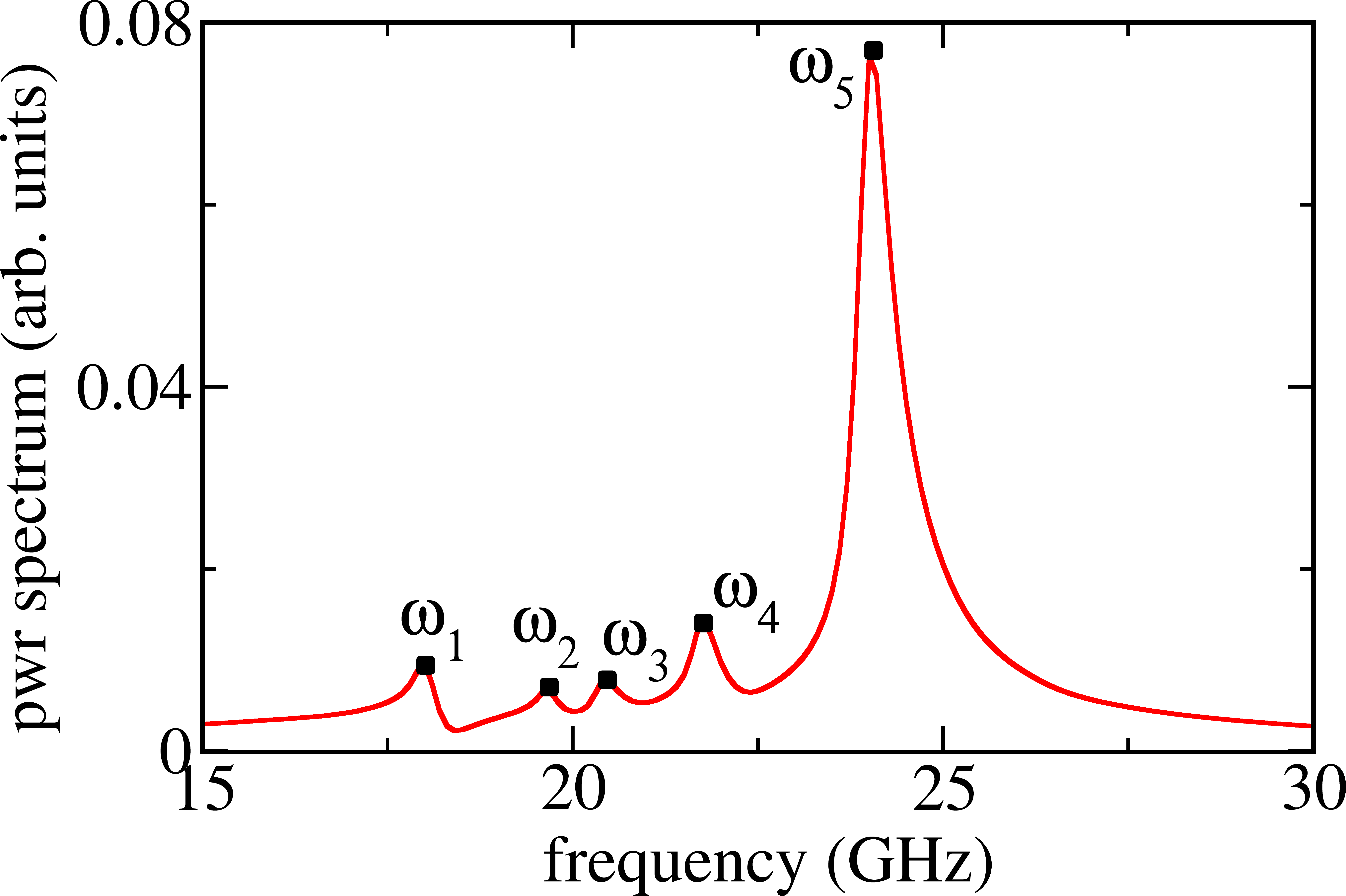}
\end{center}
\caption{SW spectrum of the system at zero temperature, characterised by 5 dipolar modes.}
\label{fig:figure3}
\end{figure}

To have a basis for comparison, we consider first dynamics at zero temperature. Starting with the magnetization uniformly tilted 5 with respect to the z axis, the time evolution is computed for 30 ns.
The SW power spectrum, shown in Fig.\ref{fig:figure3}, is given by the absolute value of the Fourier transform of the collective variable $\Psi(t)=\sum_{n=1}^{10}\psi_n(t)$.
The spectrum consists of five dipolar modes with frequencies $(\omega_1,...,\omega_5)$ of respectively $(18,19.7,20.5,21.8,24)$ GHz. By inspecting the spectra of the individual disks, 
one can see that the mode $\omega_1$ corresponds to the precession of the first and tenth disks, the mode $\omega_2$ to the second and ninth disks and so on until the fifth disk which precesses with frequency $\omega_5$,
see Fig.(\ref{fig:figure1}). This  indicates that the system has a mirror symmetry around its center, due to the fact that each disk behaves as a magnetic dipole. Aligning those dipoles in a chain gives a 
structure where the intensity of the dipolar field, which controls the frequencies, is symmetric around the center of the chain.

Let us now discuss the off equilibrium dynamics.
We consider the configuration normally adopted to study heat transfer in chains of nonlinear oscillators \cite{lepri03}, where the thermal baths act only at the boundaries of the system. 
This setup is different from the usual micromagnetic studies of the spin-Seebeck effect \cite{ohe11,hinzke11,ritzmann14,etesami14}, where each spin is coupled to a thermal reservoir with a different temperature.
The main advantage of our choice consists in that it allows to observe the spontaneous thermalisation of the system, by probing the XY temperature defined in Eq.(\ref{eq:Txy}), for the spins that \emph{are not directly connected} 
to the thermal baths. 
Experimentally, this could be realised by separating the disks with thermal insulating spacers, so that energy flows are carried only by the dipolar coupling.

The simulations at finite temperature were performed starting with the magnetisation aligned along $\hat{\bm{z}}$
and evolving the system in the presence of the thermal baths for 110 ns. The relevant observables were computed after an interval of 70 ns, necessary for the system to reach the nonequilibrium steady state.
The results were time averaged over the last 40 ns and then ensemble averaged over 32 samples with different realisation of the thermal noise.  

The high temperature bath $T_+$  ranges between 5 and 30 K, while the lower temperature bath $T_-$ is kept fix at 5 K, so that 
$\Delta T=T_+-T_-$ is comprised between 0 and 25 K. The local currents are always expressed per unit coupling and are thus pure numbers. 
According to our convention, $j_n^{M/E}$ refers to the current propagating from disk $n$ to disk $n+1$ and positive (resp. negative) currents 
propagate from left to right (resp. from right to left). 

\begin{figure}[t]
\begin{center}
\includegraphics[width=8cm]{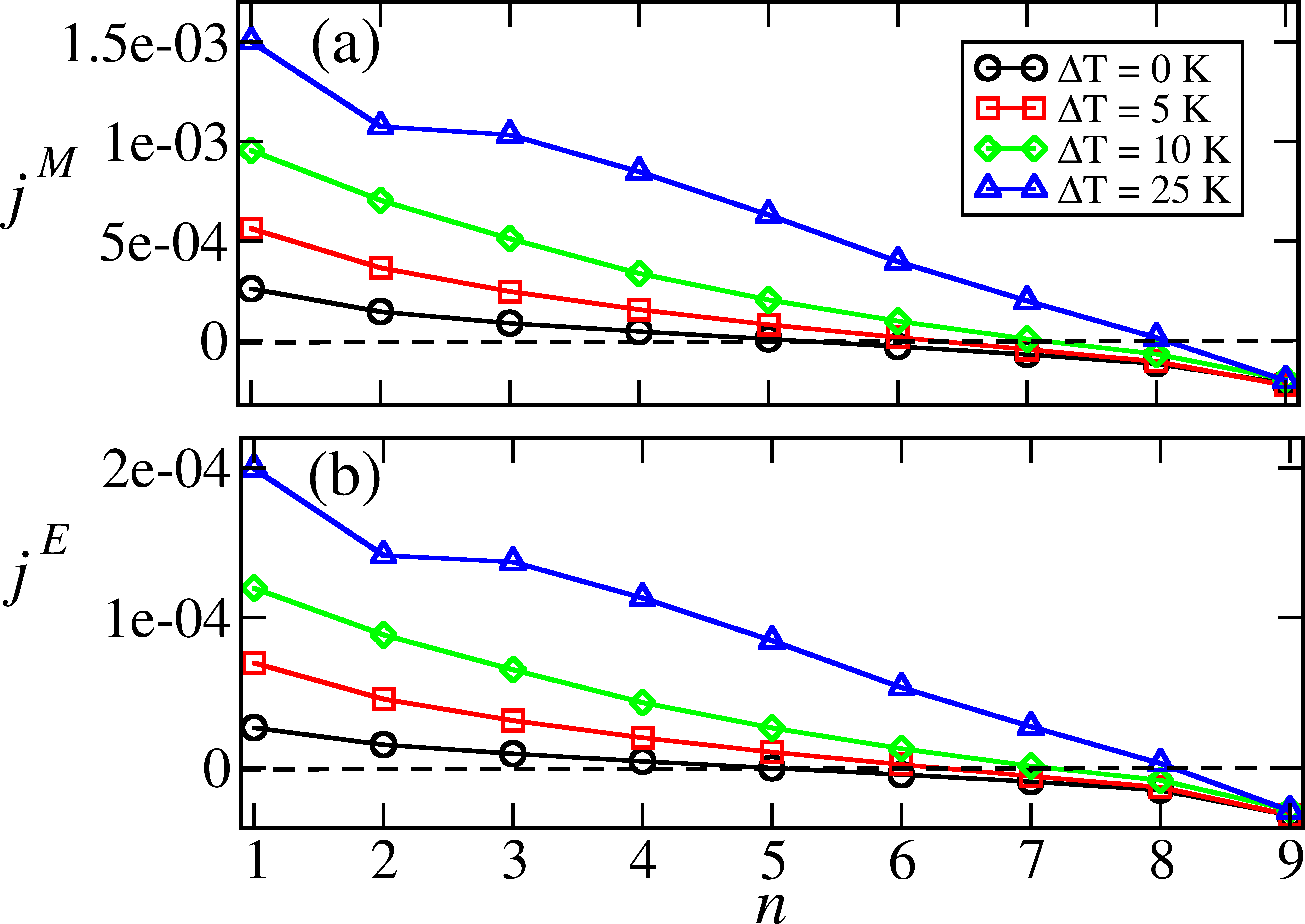}
\end{center}
\caption{Profiles of magnetisation (a) and energy (b) currents computed for different values of $\Delta T$.
The profiles are symmetric for $\Delta T=0$, and then become strongly asymmetric. The lines are guide to the eye.}
\label{fig:figure4}
\end{figure}

\begin{figure}[b]
\begin{center}
\includegraphics[width=8cm]{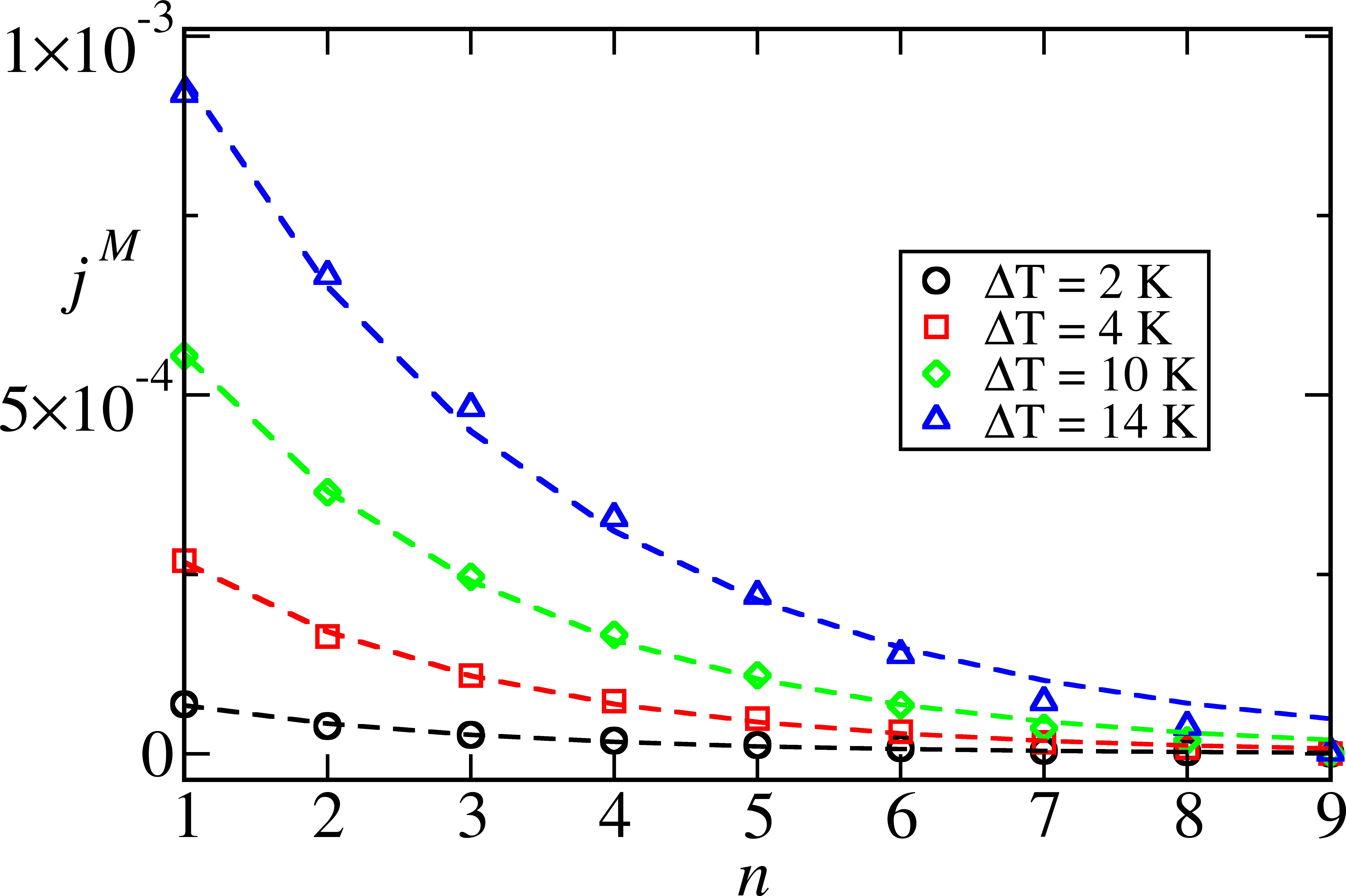}
\end{center}
\caption{Profiles of the magnetisation currents for $T_-=0$ K and different values of $T_+$. The right-going current decreases exponentially along the chain. The dashed lines are fit with $A{\rm{e}}^{-x/x_0}$,
for different values of the parameters $(A,x_0)$. The energy currents have similar profiles (not reported).}
\label{fig:figure5}
\end{figure}

\begin{figure}[t]
\begin{center}
\includegraphics[width=8cm]{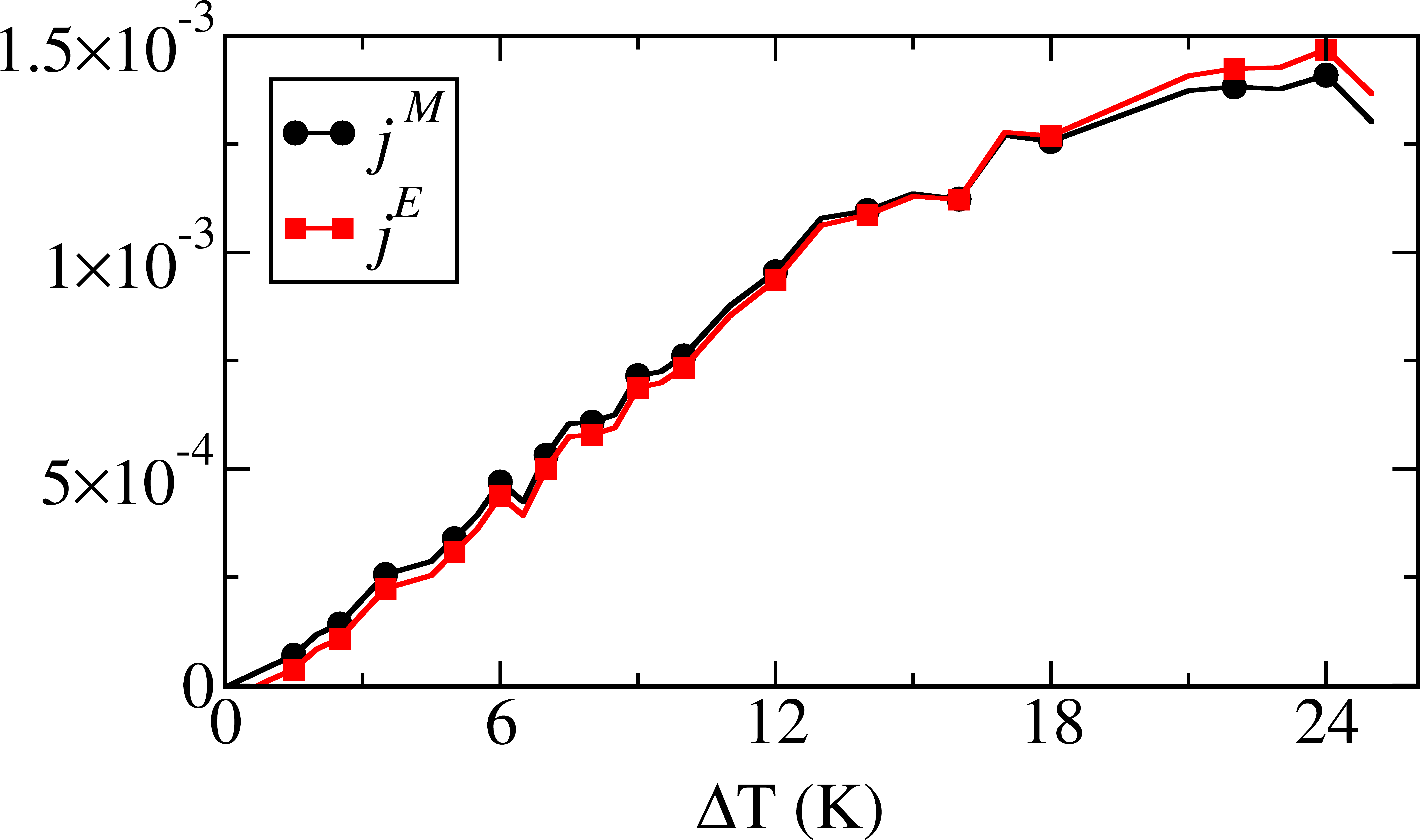}
\end{center}
\caption{Currents $j^{M/E}=j_1^{M/E}+j_9^{M/E}$ injected from the reservoirs vs the temperature difference $\Delta T$. The currents grow linearly with $\Delta T$ in the low temperature regime, 
and then reach a plateau. The lines are guides to the eye.}
\label{fig:figure6}
\end{figure}

\begin{figure}[b]
\begin{center}
\includegraphics[width=8cm]{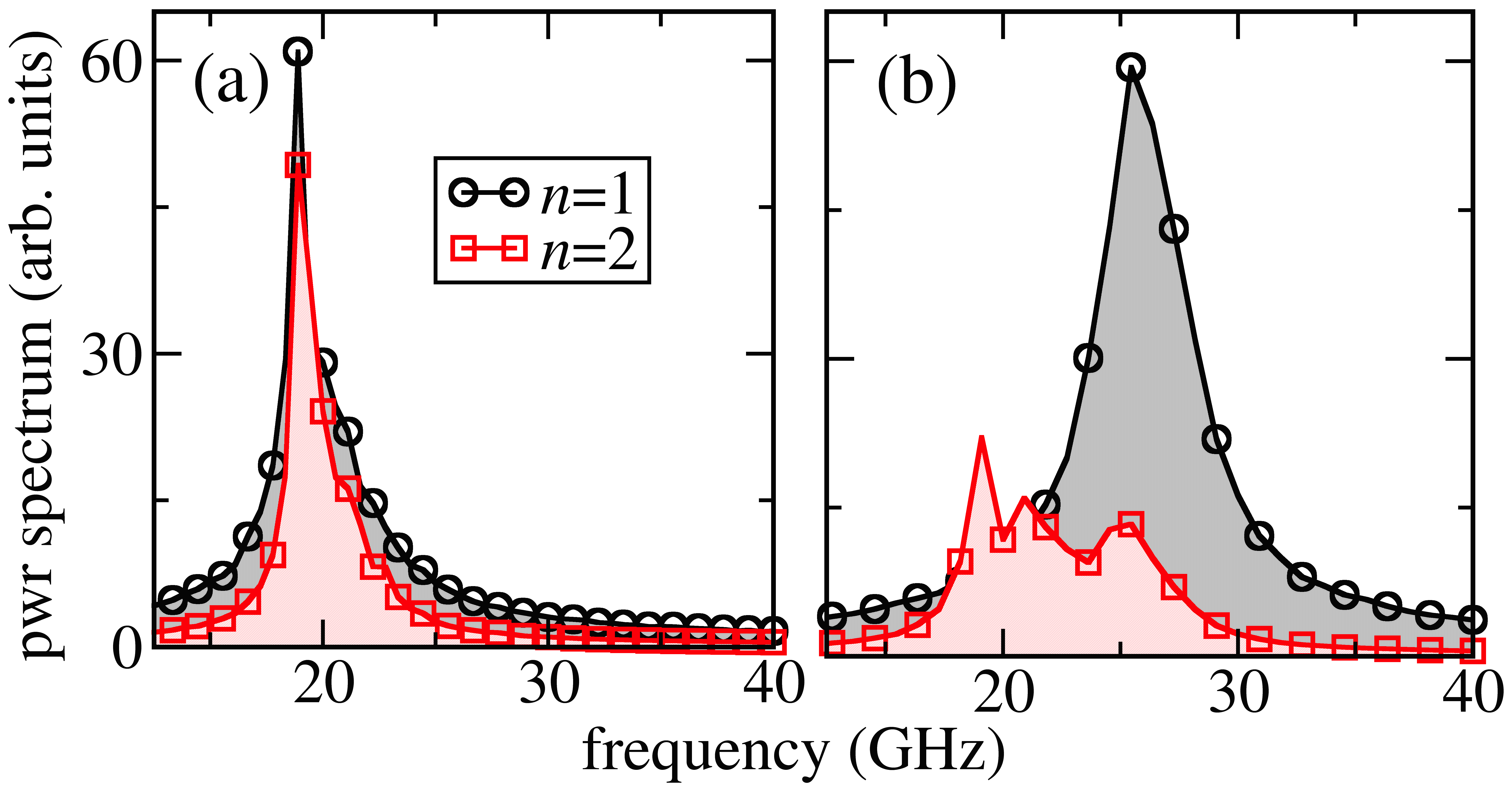}
\end{center}
\caption{Finite temperature power spectra of disks 1 and 2. 
(a) At low temperatures ($T\pm=5$ K) regime the peaks overlap and the currents are proportional to $\Delta T$. (b) At high temperature ($T_+=30$ K)
the spectrum of disk 1 shifts towards high frequency, reducing its overlap with disk 2. In this condition, the current does not increase anymore with $\Delta T$.}
\label{fig:figure7}
\end{figure}

In contrast to previous studies of the off-equilibrium DNLS, here the dissipation is both at the edges and in the bulk of the system. 
As a consequence, the currents do not have a flat profile in the bulk, but decrease exponentially along the chain.
In the thermodynamic limit, this system is thus an insulator.

Figs.\ref{fig:figure4} (a) and (b) show respectively the profiles of magnetisation and energy currents, for different values of $\Delta T$.
The two currents have similar profiles, and they do not vanish when $\Delta T=0$ (black dots). In both cases, they are symmetric with respect to 
the zero-current axis.
This behaviour is due to the fact that the currents generated by the two baths travel in opposite directions and decrease because of the damping, vanishng in the middle of the chain.
Note that, although the local currents do not vanish at equilibrium, the total current is zero.
Increasing $T_+$ leads to the increase of the right-going currents, and moves the point where the local currents vanish. 

In all these cases, the system behaves essentially as sink: the currents injected from the baths are dissipated in the bulk, so that there is no net transport. 
A situation where transport occurs can be observed in Fig.(\ref{fig:figure5}), which shows the case where $T_-=0$ and (the right-going) current is injected only by $T_+$.
The current decreases exponentially, and for high enough value of $\Delta T$ remains positive until the end of the system. Note that in the thermodynamic limit this system is an insulator, due to the damping in the bulk.

Note that the local currents do not grow indefinitely with $\Delta T$, but they saturates around $\Delta T\approx20$ K.
This can be seen more clearly in Fig.\ref{fig:figure6}, where the currents injected from the reservoirs $j^{M/E}=j_1^{M/E}+j_9^{M/E}$ grow linearly with $\Delta T$ until they reach a plateau for $\Delta T\approx 20$ K. 

This phenomenon originates from the fact that the LLG equation is nonlinear, and consequently the frequency spectrum of the system is temperature dependent.
Fig. \ref{fig:figure7} shows the power spectra of the disks 1 and 2. Panel (a) displays the low temperature regime, with $T_+=T_-=5$ K.
Thermal fluctuations excite all the modes of the system, and both disks display a broad peak between 17 and 25 GHz. The current increases linearly 
with $\Delta T$ as far those peaks overlap. Panels (b) show the high temperature case, where disk 1, directly connected to the hot bath, increases its frequency and does not overlap with disk 2. 
This situation is a manifestation of stochastic phase synchronisation (that is, the control of synchronisation through temperature) in the propagation of heat current.
A similar mechanism is at the basis of spin and thermal rectifiers \cite{casati04,ren13,ren14,borlenghi14a,borlenghi14b}.

\begin{figure}[t]
\begin{center}
\includegraphics[width=8cm]{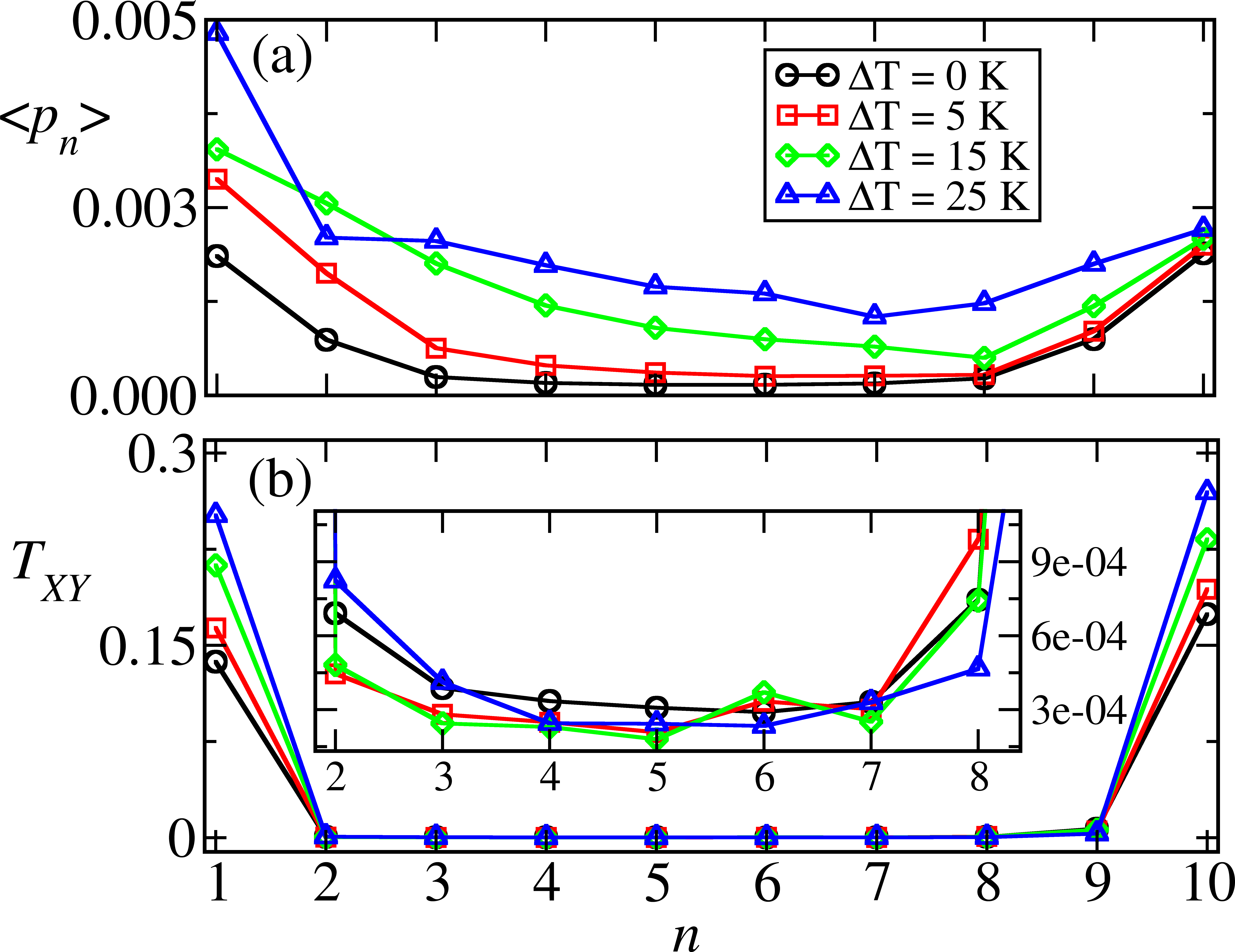}
\end{center}
\caption{(a) Profiles of SW powers $p_n$ for different values of $\Delta T$.  (b) Profiles of spin temperature $T_{XY}$ for different values of $\Delta T$.
The inset displays a magnification of the profiles between the disks 2 and 8, which show the spontaneous thermalisation of the chain. The lines are guide to the eye.}
\label{fig:figure8}
\end{figure}

The profile of the local SW powers $p_n$ is displayed in Fig.\ref{fig:figure8} (a). 
The powers reach the maximum at the edges, directly connected to the baths, and decay along the chain.
As expected, their profile is symmetric for $\Delta T=0$, and becomes strongly asymmetric when the temperature difference is
finite. Note that the powers increase with with temperature until $\Delta T\approx 15$ K and then remain roughly constant.
In this high temperature regime, $p_1$ keeps increasing because of thermal fluctuations, but $p_2$ decreases. This is due to the fact that energy remains
confined in the first disk due to the de-synchronization, as previously discussed.  

Fig.\ref{fig:figure8}(b) shows the spin temperature profiles $T_{XY}$ of the chain, computed for different values of $\Delta T$. The temperature is higher at the boundaries, where phase fluctuations are given 
by the direct contact with the baths, and then decreases dramatically towards the center of the chain, as can be seen from the inset.

\section{comparison with the oscillator model}%

\begin{figure}[t]
\begin{center}
\includegraphics[width=8cm]{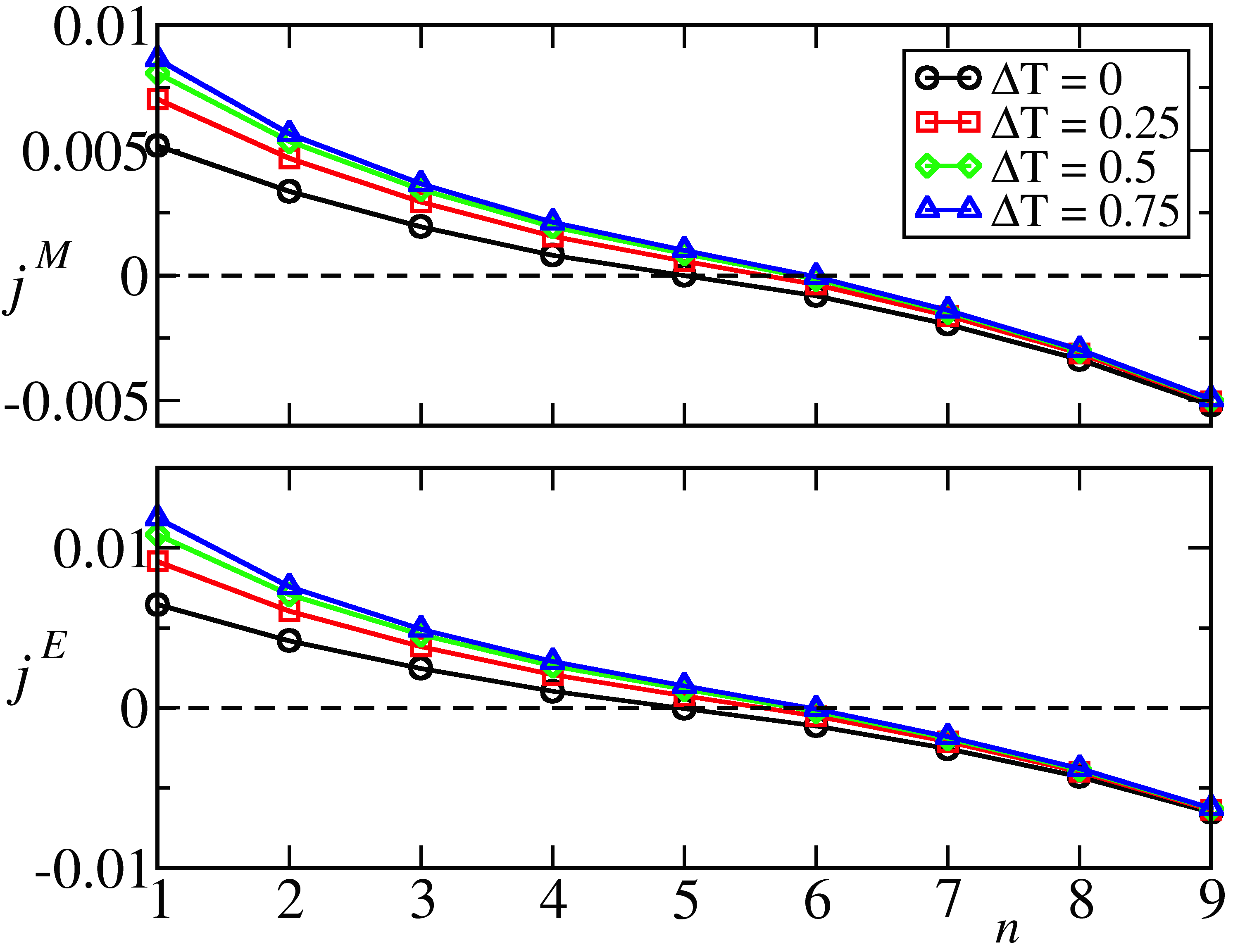}
\end{center}
\caption{Simulation of Eq.~(\ref{eq:ourdnls}) with 
$J=0.1$, $\alpha=0.008$, $\nu=1$,  $(\omega_1,\omega_2,\omega_3,\omega_4,\omega_5)
=(1,1.09,1.15,1.21,1.33)$ and different applied temperature  differences $\Delta T$ . 
The temperature of the leftmost reservoir spaces in the interval
$T_+=[0.5\,,\,1.25]$, while the rightmost one is kept fixed at $T_-=0.5$.
}
\label{fig:figure9}
\end{figure}

\begin{figure}[b]
\begin{center}
\includegraphics[width=8cm]{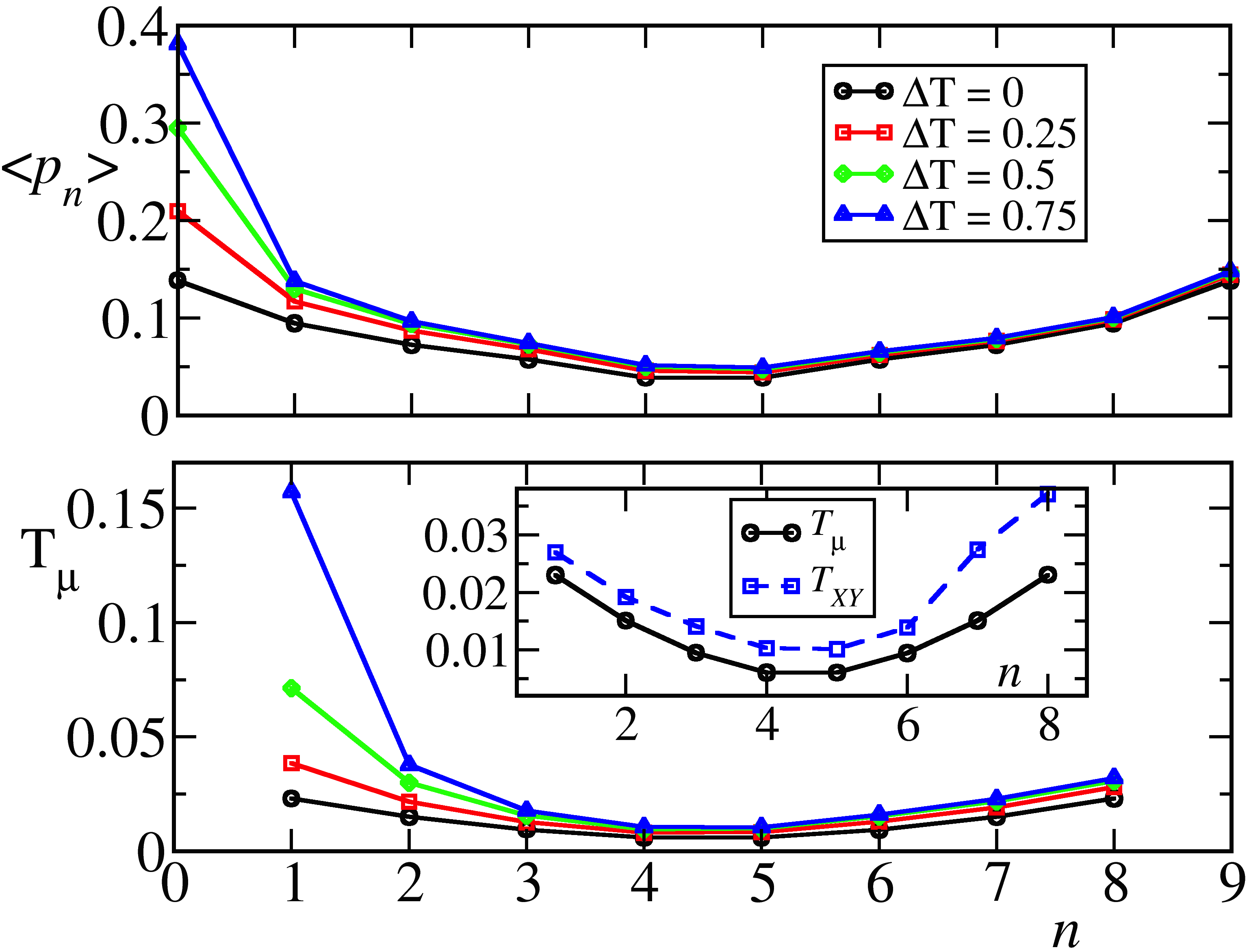}
\end{center}
\caption{Simulation of Eq.~(\ref{eq:ourdnls}), SW profiles 
and microcanonical temperature $T_\mu$ for different applied temperature differences
$\Delta T$.
The parameters of the simulation are the same as in Fig. \ref{fig:figure9}. 
For each lattice site $n$, the local temperature $T_\mu$ is calculated by evaluating the general 
definition~\cite{iubini12,iubini13} on a triplet of sites
centered around $n$. The inset of the lower panel shows a comparison between $T_\mu$ and $T_{XY}$ for
the profile with $\Delta T=0$.
}
\label{fig:figure10}
\end{figure}

The micromagnetic simulations indicate that the system has some form of mirror symmetry around its center. Thus we model this by choosing
the linear frequencies  in model (\ref{eq:ourdnls}) such that $\omega_n^0=\omega_{N-n}^0$. A relatively small coupling $J=0.1$
has been employed. As in Section \ref{sec:MMag}, the fluctuating forces are applied  only at the boundaries of the system, i.e. we choose $T_n = [T_+\delta_{n,1}+T_-\delta_{n,N}]$; 

As a test for stationarity we evaluated the average currents from the heat baths $ \sum_n\langle\mathcal{J}_n^M\rangle$ and 
$ \sum_n \langle\mathcal{J}_n^H\rangle$ are indeed vanishing within statistical accuracy.

In Figs. \ref{fig:figure9} and \ref{fig:figure10} we show the current and SW power profiles along the chain.
Comparing with the
corresponding figures \ref{fig:figure4}(a,b) and \ref{fig:figure8}(a), the qualitative agreement is good.
The lower panel of Fig. \ref{fig:figure10} reports the temperature profiles $T_\mu$ calculated
with the exact microcanonical expression defined in Refs.\cite{iubini12,iubini13}. Such profiles are in good agreement with the phase temperature
$T_{XY}$ (see the inset) and qualitatively reproduce
the temperature profiles calculated within the micromagnetic  framework in Fig. \ref{fig:figure2}.
Altogether, this confirms that definition (\ref{eq:Txy}) 
is a sensible approximation in the present setup.   

\begin{figure}[t]
\begin{center}
\includegraphics[width=8cm]{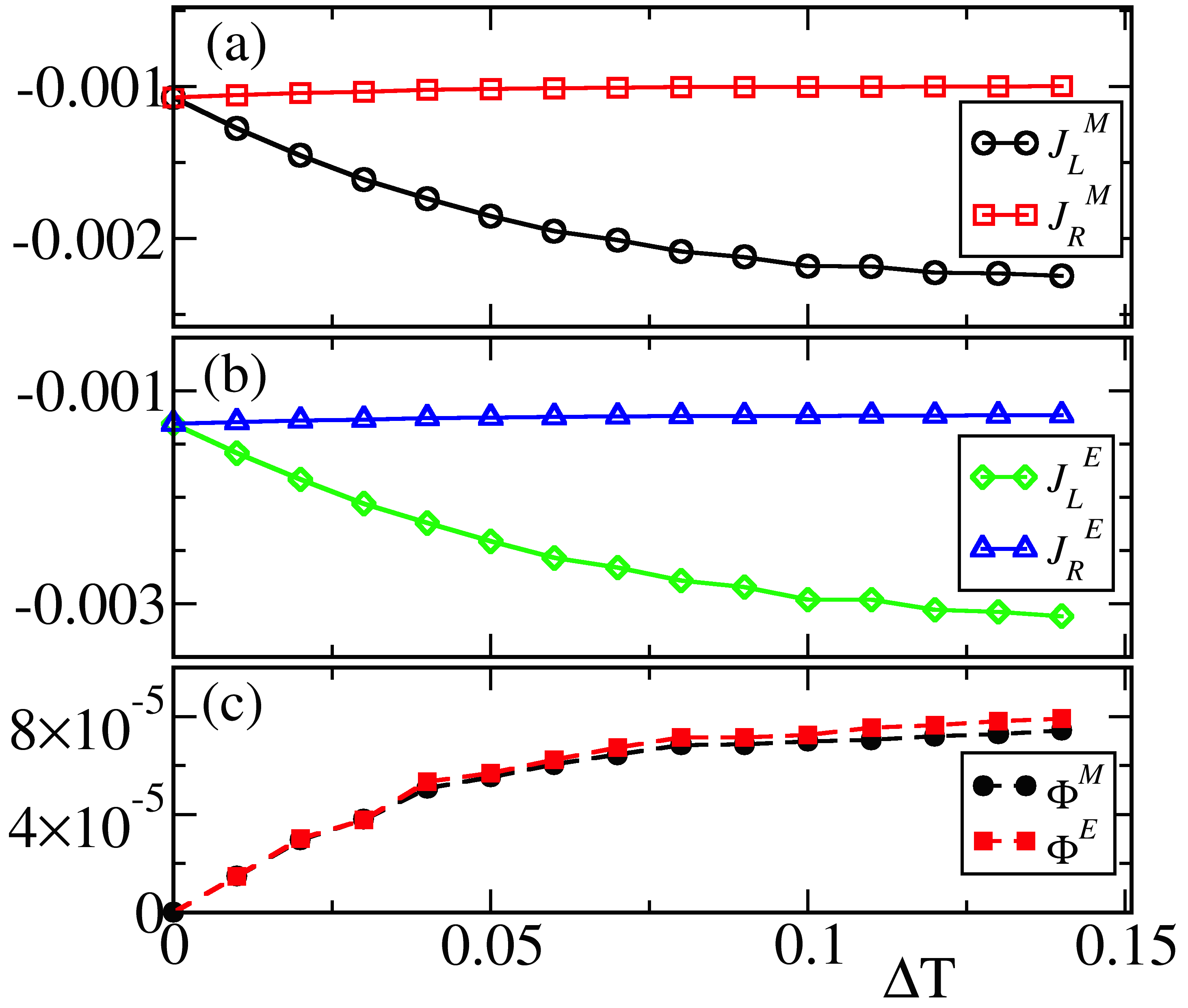}
\end{center}
\caption{Simulation of Eq.~(\ref{eq:ourdnls})
Boundary magnetisation (a) and energy (b) currents toward the two left and right reservoirs 
versus the temperature difference $\Delta T$
for $T_-=0.05$, other parameters given in the text. Panel (c) shows the corresponding excess 
fluxes calculated with Eq.(\ref{eq:eff_flux}).}
\label{fig:figure11}
\end{figure}

Finally, in panels (a) and (b) of Fig.\ref{fig:figure11}  we show the stationary boundary fluxes
$\mathcal{J}^M_n$ and $\mathcal{J}^E_n$.
Nicely, they reproduce the   saturation effect observed
in Fig.\ref{fig:figure4} with 
the micromagnetic simulations. It should be noted that the currents do not vanish for $\Delta T=0$: this  is 
simply because the system is off-equilibrium also in this case 
due to the presence of the dissipation in the bulk of the chain. For the very same reason, the two currents
flowing at the boundary are 
not equal.  In order to single out the purely transport contribution from the dissipative one in the boundary
fluxes, we  define an {\it excess} boundary flux $\Phi^{M,E}$ as

\begin{equation}\label{eq:eff_flux}
\Phi^{M,E}= \mathcal{J}_R^{M,E}(\Delta T) - \mathcal{J}_R^{M,E}(0) \quad,
\end{equation}
that takes into account the amount of magnetisation/energy which is transported per unit time with respect to the
purely dissipative symmetric profile  at $\Delta T=0$. The behaviour of $\Phi^{M,E} (\Delta T)$ 
is shown in Fig.\ref{fig:figure11}(c). Both the excess fluxes have positive sign, meaning that there 
is a net transport of energy and magnetisation from left to right. Moreover they show the same saturation
effect observed in panels (a) and (b).

\section{conclusions}%

In this work we have illustrated the  spin-Seebeck effect in a small array of 
coupled magnetic nanodisks and we have  related to general transport properties of out-of-equilibrium chains of nonlinear oscillators.
The dynamics of the disk chain has been investigated by means of micromagnetic simulations
and compared with the effective model, Eq.~(\ref{eq:ourdnls}). There is indeed 
a good qualitative agreement between the two approaches, and a good physical
insight can be achieved from the coupled oscillator model.

Another remarkable result, is that the relationship with the XY model provides a 
simple prescription for computing the local phase temperature in the micromagnetic simulations
via Eq.~(\ref{eq:Txy}). Such observable plays a relevant role in our setup, since 
it allows to quantify thermal fluctuations inside the system, i.e. for macrospins that
are not directly connected to the external reservoirs.
This last issue is of major 
importance for non-standard Hamiltonians like the DNLS one, where kinetic and 
potential energies are not separated. Indeed, the XY approximation 
allows introducing the simple kinetic expression $T_{XY}$ for the temperature,
that can safely approximate the microcanonical one $T_\mu$.
This is of practical importance, considering that the microscopic 
definitions of $T$ and $\mu$ are pretty much involved for a
non separable Hamiltonian, like the DNLS one.

The presence of two coupled currents is related to the existence of two thermodynamic forces
\cite{onsager31a,onsager31b,kubo83} $(\Delta T,\Delta\mu)$, the latter being the difference of chemical potential \cite{iubini12,iubini13}. In this work we limited to the case where $\Delta\mu=0$. Actually, chemical potential gradients 
can be easily accounted for within the DNLS language, 
by adding terms of the form $i\alpha \mu_n\psi_n$ to Eq.\ref{eq:ourdnls} \cite{iubini13}.
For our system of precessing spins, this is interpreted
as a torque that compensates the damping 
and controls the magnon relaxation time towards the reservoirs \cite{iubini12,iubini13,borlenghi14b}. 
This can be experimentally realised trough spin transfer torque \cite{slonczewski96,berger96,slavin09}. 
In this case the spin-torque induces a nonequilibrium dynamics and may 
lead to self-sustained oscillations, thus opening a new realm of transport 
phenomena. We plan to investigate those setups in the next future.

\section*{Acknowledgements}%

We thank Prof. Magnus Johansson for illuminating discussions. We acknowledge financial support from the Swedish Research Council (VR), Energimyndigheten (STEM), the Knut and Alice Wallenberg Foundation, the Carl Tryggers Foundation, the Swedish e-Science Research Centre (SeRC) and the Swedish Foundation for Strategic Research (SSF). S.I. acknowledges financial support from the EU-FP7 project PAPETS (GA 323901). We gratefully acknowledge the hospitality of the Galileo Galilei Institute for Theoretical Physics, where part of this work was performed, during the 2014 workshop \emph{Advances in Nonequilibrium Statistical mechanics}.


\end{document}